\documentstyle[prd,aps,amstex,epsf,preprint]{revtex}
\tightenlines
\newcommand{\bea}{\begin{eqnarray}}  \newcommand{\eea}{\end{eqnarray}}
\newcommand{\beq}{\begin{equation}}  \newcommand{\eeq}{\end{equation}}
  
\newcommand{\lmk}{\left(}  \newcommand{\rmk}{\right)}
\newcommand{\lnk}{\left\{}  \newcommand{\rnk}{\right\}}
\newcommand{\lkk}{\left[}  \newcommand{\rkk}{\right]}

\newcommand{\vect}[1]{\mbox{\boldmath${#1}$}}

\newcommand{\ev}{{\rm ev}}
\newcommand{\tot}{{\rm tot}}
\newcommand{\crr}{{\rm cr}}
\newcommand{\hc}{{\rm H.C.}}
\newcommand{\Ree}{{\rm Re}}
\newcommand{\Imm}{{\rm Im}}
\newcommand{\os}{{\rm os}}
\begin{document}

\preprint{YITP-02-61, OU-TAP 185, RESCEU-16/02}

\title{Stabilizing dilaton and baryogenesis}

\author{Alexander D. Dolgov}
\address{Yukawa Institute for Theoretical Physics, Kyoto
University, Kyoto 606-8502, Japan,\\
INFN, sezione di Ferrara, via Paradiso, 12, 44100 - Ferrara,
Italy, \\
and
ITEP, Bol. Cheremushkinskaya 25, Moscow 113259, Russia.}

\author{Kazunori Kohri}
\address{Research Center for the Early Universe, University of Tokyo, 
Tokyo
113-0033, Japan } 

\author{Osamu Seto}
\address{Yukawa Institute for Theoretical Physics, Kyoto
University, Kyoto 606-8502, Japan }

\author{Jun'ichi Yokoyama}
\address{Department
of Earth and Space Science, Graduate School of Science,\\ Osaka
University, Toyonaka 560-0043, Japan }

\maketitle

\begin{abstract} 
Entropy production by the dilaton decay is studied in the model
where the dilaton acquires  potential via gaugino condensation in 
the hidden gauge group.  Its effect on the 
Affleck-Dine baryogenesis is investigated with
and without non-renormalizable terms in the potential. It 
is shown that the baryon asymmetry produced by this
mechanism with the higher-dimensional terms is diluted by the
dilaton decay and can be regulated to the observed value.
\end{abstract}

\section{introduction}

Many gauge-singlet scalar fields arise in the effective four-dimensional
supergravity which could be derived from string theories.  Among them 
the dilaton $S$ has a flat potential in all orders in perturbation
theory \cite{Dine:1986vd}.
Therefore some non-perturbative effects are expected to
generate the potential whose minimum corresponds to the vacuum
expectation value (VEV). The most promising mechanism of the dilaton
stabilization and the supersymmetry breaking is the gaugino
condensation in the hidden gauge sector
\cite{Dine:1985rz,Derendinger:1985kk,Casas:1990jy,deCarlos:1992da}.

In cosmological consideration, even if the dilaton acquires 
potential through such non-perturbative effects, there are some
difficulties to relax the dilaton to the correct minimum, as pointed out
by Brustein and Steinhardt \cite{Brustein:1993nk}.  Since the
potential generated by multiple gaugino condensations is very steep in 
the small field-value region, the dilaton would have large kinetic
energy and overshoot the potential maximum to run away to infinity. 

As a possible way to overcome this problem, Barreiro \textit{et al.}
\cite{Barreiro:1998aj} pointed out that the dilaton could slowly roll
down to its minimum with a little kinetic energy due to large Hubble
friction if the
background fluid dominated the cosmic energy density.
As a result, the dilaton could be trapped and would oscillate around the
minimum.

When the dilaton decays, however, the remaining energy density is
transformed to radiation.  One may worry about a huge entropy production
by the decay, because it
could dilute initial baryon asymmetry \cite{Coughlan:1983ci}. 
Therefore in order to
obtain the observed baryon asymmetry, as required by {\it e.g.} 
nucleosynthesis, it is
necessary to produce a larger asymmetry than the observed one at the 
outset.

The attractive mechanism to produce large baryon asymmetry in
supersymmetric models was proposed by Affleck and Dine
\cite{Affleck:1984fy}. However, the baryon-to-entropy ratio  produced
by this mechanism, $n_b/s$, is usually too large.  So if  we take into
account the entropy production after the baryogenesis,
we can expect that the additional entropy may dilute the
excessive baryon asymmetry to the observed value, as pointed out in {\it
e.g.}
\cite{Linde:1985jy,Campbell:1998yi}.

In this paper, we investigate whether the dilution by 
the dilaton decay can regulate the large baryon asymmetry produced by 
the
Affleck-Dine mechanism to the observed value. 
The  paper is organized as
follows. In  \S II and \S III, we describe the potential and the 
dynamics
of the dilaton. Then, in \S IV
 we estimate the baryon asymmetry generated by the 
Affleck-Dine mechanism taking into account dilution by the
dilaton decay. Section V is devoted to the conclusion.
We take units with $8\pi G = 1$.

\section{dilaton potential}

We consider the potential of the dilaton non-perturbatively induced by
multiple gaugino condensates. In string models, the tree level
K\"{a}hler potential is given by
\begin{equation}
K = -\ln(S+S^*)-3\ln(T+T^*-|\Phi|^2) , \label{kahler} 
\end{equation}
where $S$ is the dilaton, $T$ is the modulus and $\Phi$ represents some
chiral matter fields \cite{Witten:1985xb}.

The effective superpotential of the dilaton \cite{Dine:1985rz}
generated by the gaugino condensation is given by
\begin{equation}
W = \sum_a \Lambda_a(T) e^{-\alpha_a S} .
\end{equation}
Here, for the $SU(N_a)$ gauge group and the chiral matter in
$M_a(N_a+\bar{N}_a)$ ``quark'' representations, $\alpha_a$ and
$\Lambda_a$ are given by
\begin{equation}
\alpha_a = \frac{8\pi^2}{N_a-\frac{M_a}{3}} ,
\end{equation}
\begin{equation}
\Lambda_a = -\frac{N_a-\frac{M_a}{3}}{\eta^6(T)}(32\pi^2 e)^{3(N_a-M_a/3)/(3N_a-M_a)}\left(\frac{M_a}{3}\right)^{M_a/(3N_a-M_a)} .
\end{equation}
Then the modulus $T$ have a potential minimum due to the presence of 
the Dedekind function $\eta(T)$ \cite{deCarlos:1992da}. 
Hereafter we assume the stabilization of the modulus $T$ at the
potential minimum and concentrate on the evolution of the real part of
the dilaton field.

Since at least two condensates are required to form the potential minimum,
we consider a model with two condensates. Then the indices of the gauge
group are $a = 1,2$, and we take $10 \lesssim \alpha_1 \lesssim \alpha_2
$. 
 
The potential $V$ for scalar components in supergravity is
given by 
\begin{equation}
V = e^K\left[(K^{-1})^i{}_jD_i W(D_j W)^*-3|W|^2\right]  ,\label{pot}
\end{equation}
where
\begin{equation}
D_i W = \frac{\partial W}{\partial\Phi^i}+\frac{\partial
K}{\partial\Phi^i}W ,\label{deri}
\end{equation}
$K_i{}^j = \partial^2 K/\partial\Phi^i\partial\Phi_j^*$, the inverse 
$(K^{-1})^i{}_j$ is defined by $(K^{-1})^i{}_jK^j{}_k = \delta^i{}_k$
and $i = S, T, \Phi$.
In the region, $\alpha \Ree S \gg 1$, the potential Eq.\ (\ref{pot})
can be rewritten as
\begin{eqnarray}
V(S) &\simeq& e^K(K^{-1})^S{}_S|\partial_S W|^2 \nonumber \\ 
  & = & (S+S^*)|\alpha_1\Lambda_1|^2
  e^{-\alpha_1(S+S^*)} \left|1+\frac{\alpha_2\Lambda_2}{\alpha_1\Lambda_
1}e^{-(\alpha_2-\alpha_1)S}\right|^2 \label{pot2}.
\end{eqnarray}
In this potential the imaginary part of $S$ has a minimum at
\begin{eqnarray}
\Imm S_{\min} = \frac{(2n+1)\pi}{\alpha_1-\alpha_2},  
\end{eqnarray}
where $n $ is an integer.
So we assume $\Imm S= \Imm S_{\min} $ and concentrate
on the behavior of $\Ree S$ hereafter. Then we find that the potential
minimum, $\Ree S_{\min}$, is given by
\begin{equation}
\Ree S_{\min} =
\frac{1}{\alpha_2-\alpha_1}\ln\left(\frac{\alpha_2}{\alpha_1}
\frac{\Lambda_2}{\Lambda_1}\right) .
\end{equation}
We assume $\Ree S_{\min}\simeq 2$ to reproduce a 
phenomenologically viable value of the gauge coupling constant of
grand unified theory \cite{deCarlos:1992da}.
The negative vacuum energy at the minimum of the potential 
is assumed to be canceled by some mechanism such as a vacuum expectation 
value of three form field strength.
Although we consider a model with two gaugino condensates, 
our following estimations would be almost unchanged in a single
condensate model with non-perturbative K\"{a}hler corrections
\cite{Casas:1996zi,Binetruy:1997vr}, because the evolution of the
dilaton is determined only by the slope of the potential in the 
region $S \ll S_{\min}$ and its mass.

On the other hand, the position of the
local maximum of potential, $\Ree S_{\max}$, is given 
by
\begin{equation}
\Ree S_{\max} = \Ree S_{\min} + \frac{1}{\alpha_2-\alpha_1}\ln\left(\frac{\alpha_2
}{\alpha_1}\right) .
\end{equation}

For $S \ll S_{\min}$, the potential (\ref{pot2}) can be
approximated as
\begin{equation}
V(S) \simeq 2 \Ree S V_0e^{-\alpha_2 2\Ree S} \label{pot3},
\end{equation}
where $V_0 \simeq |\alpha_2\Lambda_2|^2$. The potential (\ref{pot2}) has the
minimum at $S_{\min}$ and the local maximum at $S_{\max}( > S_{\min})$.
For $S \gtrsim S_{\crr} \equiv S_{\min}-1/(\alpha_2-\alpha_1)$,
the approximate expression for the potential (\ref{pot3}) breaks down. 
Around the
potential minimum $S_{\min}$, the potential (\ref{pot2}) becomes
\begin{equation}
V(S) \simeq |\alpha_1\Lambda_1|^2 e^{-2\alpha_1
S_{\min}}(\alpha_2-\alpha_1)^2(\Ree S -\Ree S_{\min})^2 .
\end{equation}

However, one can see from 
the K\"{a}hler potential (\ref{kahler}) that the variable
$\textrm{Re} S$ does not have the canonical kinetic term. Therefore we
introduce the canonically normalized variable, $\phi$, as
\begin{equation}
\phi \equiv \frac{1}{\sqrt{2}}\ln\textrm{Re} S.
\end{equation}

\section{stabilization mechanism for the dilaton}

Here, after reviewing the mechanism for dilaton stabilization
proposed by Barreiro \textit{et al.} \cite{Barreiro:1998aj}, we
estimate the relic energy density of the dilaton and the amount of the 
entropy density produced by its decay. We will consider the situation
that the universe after inflation contains the dilaton, $\phi$, and a
fluid with the equation of state, $p = (\gamma-1)\rho$, where $\gamma$
is a constant. For example, $\gamma = 4/3$ for radiation or $\gamma = 1$
for non-relativistic matter.  The latter includes oscillating 
inflaton field or/and the Affleck-Dine (AD) condensate $\phi_{AD}$.

In the spatially flat Robertson-Walker space-time,
\begin{equation}
ds^2 = -dt^2+a(t)^2 d\vect{x}^2 ,
\end{equation}
with the scale factor $a(t)$, the Friedmann equations and the field
equation for $\phi$ read
\begin{eqnarray}
\dot{H} &=& -\frac{1}{2}(\rho+p+\dot{\phi}^2) , \\
\ddot{\phi} &=& -3H\dot{\phi} - \frac{d V(\phi)}{d \phi} ,\\
H^2 &=& \frac{1}{3}\lkk\rho+\frac{1}{2}\dot{\phi}^2+V(\phi)\rkk , 
\end{eqnarray}
where $H = \dot{a}/a$ is the Hubble parameter and a dot denotes time
differentiation. We define the new variables,
\begin{equation}
x \equiv \frac{\dot{\phi}}{\sqrt{6}H},\ 
~~~~~~~~~~\ y \equiv \frac{\sqrt{V(\phi)}}{\sqrt{3}H},
\end{equation}
and the number of $e$-folds $N \equiv \ln(a)$.

Then, the equations of motion can be rewritten as 
\begin{eqnarray}
x' &=& -3x-\sqrt{\frac{3}{2}}\frac{\partial_{\phi}V}{V}y^2
+\frac{3}{2}x\lkk 2x^2+\gamma(1-x^2-y^2)\rkk , \label{eomx} \\
y' &=& \sqrt{\frac{3}{2}}\frac{\partial_{\phi}V}{V}xy 
+\frac{3}{2}y \lkk 2x^2+\gamma(1-x^2-y^2)\rkk ,  \label{eomy} \\
H' &=& -\frac{3}{2}H\lkk 2x^2+\gamma(1-x^2-y^2)\rkk  \label{eomh},
\end{eqnarray}
where the prime denotes a derivative with respect to $N$. In terms of
these variables, the Friedman equation becomes $x^2+y^2+\rho/(3H^2)
=1$. 
We see that $x^2$ and $y^2$ are respectively the ratios of the
kinetic and potential energy densities of the dilaton to the total
energy density. We consider the case of the universe  dominated by
the background fluid, so the inequalities, $x^2, y^2 \ll 1$,
hold. Then Eq.\ (\ref{eomh}) can easily be solved and the solution is 
\begin{equation}
H = H_0 e^{-3\gamma N/2} \label{h}.
\end{equation}
Next we introduce another new variable,
\begin{equation}
x_s \equiv \frac{d\text{Re}S}{d\phi}x.
\end{equation}
Then Eqs.\ (\ref{eomx}) and (\ref{eomy}) can be respectively rewritten 
as 
\begin{eqnarray}
x_s' &=& -3x_s+\frac{3}{2}\gamma x_s+2\alpha_2\sqrt{\frac{3}{2}}
\left(\frac{d\text{Re}S}{d\phi}\right)^2y^2 , \\
y' &=& -2\alpha_2\sqrt{\frac{3}{2}}x_sy +\frac{3}{2}\gamma y ,
\end{eqnarray}
 where we have used  the relation $-2\alpha_1V = \partial
V/\partial\text{Re} S$. Now we examine the stationary points in the
above equations. From $y' = 0$, we find 
\begin{equation}
x_s = \sqrt{\frac{3}{2}}\frac{\gamma}{2\alpha_2}. \label{x}
\end{equation}
On the other hand, from $x_s' =0$ we see
\begin{equation}
y^2 =
\frac{3(2-\gamma)\gamma}{2(2\alpha_2)^2}
\left(\frac{d\text{Re}S}{d\phi}\right)^{-2}. \label{y} 
\end{equation}
Except for the factor $(d\text{Re}S/d\phi)^{-2}$, Eqs.\ 
 (\ref{x}) and  (\ref{y}) represent the scaling solution for the
scalar field with exponential potential \cite{Copeland:1997et}.
In spite of the presence of the factor $(d\text{Re}S/d\phi)^{-2}$, we
can verify that the deviation from the scaling solution is small
enough, as already shown in \cite{Barreiro:1998aj}.  Thus we can write 
the
solution as
\begin{equation}
\text{Re}S =
\frac{3\gamma}{2\alpha_2}N+\frac{1}{\alpha_2}\ln\left[\frac{4V_0}{(2-
\gamma)\gamma}\left(\frac{2\alpha_2}{3H_0}\right)^2\right]+\epsilon(N) ,
\label{s}
\end{equation}
where 
\begin{equation}
\epsilon(N) =
\frac{2}{\alpha_2}\ln\lnk\frac{3\gamma}{2\alpha_2}N+\frac{1}{\alpha_2}
\ln\lkk\frac{4V_0}{(2-\gamma)\gamma}\left(\frac{2\alpha_2}{3H_0}\right)^2
\rkk\rnk ,
\end{equation}
denotes the deviation from the scaling solution;
we find $x_s' = \epsilon(N)'' \ll 1$. 

Figure 1 depicts time evolution of Re$S$ as a function of $N$ for
various initial values of the dilaton field amplitude.  
As is seen there if the dilaton
relaxes to the scaling solution before reaching $S_{\min}$, its energy
is small enough to prevent overshooting.  Attractor behavior in a
different situation has been studied in \cite{Yokoyama:1988km}.
The Hubble parameter at $S = S_{\min}$ is estimated as
\begin{equation}
H_{\min} \simeq
\frac{2\alpha_2}{3}\sqrt{\frac{2V_0}{(2-\gamma)\gamma}}
e^{-\alpha_2S_{\min}}\equiv t^{-1}_{\min},
\end{equation}
from Eqs.\ (\ref{h}) and (\ref{s}). On the other hand, as is seen from
Eq.\ (\ref{pot2}), the mass of the 
dilaton in  vacuum is given by
\begin{equation} 
m_{\phi} \simeq 2(\alpha_2-\alpha_1)\sqrt{V_0}e^{-\alpha_2 S_{\min}}.
\end{equation}
We therefore find $m_{\phi} \simeq H_{\min}$, so the dilaton 
begins to oscillate immediately when it approaches the potential
minimum. Since the
mass of the gravitino is given by $m_{3/2} \simeq \Lambda_2
e^{-\alpha_2 S_{\min}}$, the mass of the dilaton is
\begin{eqnarray}
m_{\phi} &\simeq& \alpha_1^2 m_{3/2} \nonumber \\
 &\simeq&
 10^2 \textrm{TeV}
\left(\frac{m_{3/2}}{1 \textrm{TeV}}
\right)\left(\frac{\alpha_1}{10}\right)^2 \label{mass}. 
\end{eqnarray}
When the dilaton $S$ approaches the critical point $S_{\crr}$, the
single exponential approximation (\ref{pot3}) breaks down and the 
scaling
behavior terminates. Then the energy density of the dilaton is
estimated as
\begin{eqnarray}
\rho_\phi^{in} =
\left.\frac{1}{2}\dot{\phi}^2+V \right|_{t_{\min}} &=& 3H^2\left.(x^2+
y^2)\right|_{t_{\min}} \nonumber \\ 
 &=&
 \left.\frac{3}{2}\left(\frac{\gamma}{2\alpha_1}\right)^2
\rho\left(\frac{d\text{Re}S}{d\phi}\right)^{-2}
\left(1+\frac{2-\gamma}{2\gamma}\right)
\right|_{t_{\min}} \nonumber \\ 
 &\simeq& 10^{-3}\rho\left(\frac{10}{\alpha_1}\right)^2\left(\frac{2}
{\text{Re}S}\right)^2 \ ~~~~~~~~~~\
 \text{for} \ ~~~~~~~~~~\ \gamma = 4/3 . \label{relicenergy}
\end{eqnarray}
Indeed, the numerical
calculation gives a close value
$\rho_{\phi} \simeq 10^{-4}\rho$ at the
beginning of the oscillation. After that, the dilaton begins to 
oscillate and the energy density
decreases as $a(t)^{-3}$ until it decays.

In Fig.\ 2, we present the energy density of the dilaton in the universe
filled by the $\gamma = 4/3$ background fluid at the
beginning of the oscillation regime for various model parameters,
$\alpha_1$ and $\alpha_2$. 
The values
along the contour lines represent the energy density $\rho_{\phi}$ in 
the
unit of $10^{-4}\rho$. 
 The case with
$\gamma = 1$ is depicted in Fig.\ 3, where we find smaller energy
density of the dilaton by a factor of $\sim 3$.
These figures are drawn in the two-parameter
space, though the potential contains four parameters as seen from
Eq.\ (\ref{pot2}).  The other two have been fixed by 
setting $m_{3/2} = 1$ TeV and
$S_{\min} = 2$ \cite{deCarlos:1992da}. 

The decay of the dilaton produces huge entropy. If we assume that 
the background fluid is radiation with $\gamma = 4/3$, at the time
$t = t_{\min}$ its temperature is $T \sim 10^{11}$ GeV and the
entropy density is 
\begin{equation}
s = \frac{4\pi^2}{90}g_{*}T^3 , 
\end{equation}
where $g_{*} \sim 10^2$ is the 
effective number of relativistic degrees of freedom.
By using Eq.\ (\ref{relicenergy}), we find that the entropy density 
increases 
by the factor 
\begin{eqnarray}
\Delta =
\frac{T}{T_D}\left.\frac{\rho_{\phi}}{\rho}\right|_{t_{\min}}  
 \simeq
 10^{9}\left(\frac{T}{10^{11}\textrm{GeV}}\right)\left(\frac{10^{-2}
\textrm{GeV}}{T_D}\right)
\left(\frac{\left.\rho_{\phi}/\rho\right|_{H_{\min}^{-1}}}{10^{-4}}\right) ,
\end{eqnarray}
when the dilaton decays at 
\begin{eqnarray}
H = \Gamma_D \simeq m_{\phi}^3 
 \simeq 10^{-21}\textrm{GeV}
\left(\frac{m_{\phi}}{10^2\textrm{TeV}}\right)^3,
\end{eqnarray}
where
\begin{eqnarray}
T_D \simeq m_{\phi}^{3/2} 
\simeq
10^{-1}\textrm{GeV}\left(\frac{m_{\phi}}{10^2\textrm{TeV}}\right)^{3/2} ,
\end{eqnarray}
is the reheating temperature after decay of the dilaton.

\section{Affleck-Dine baryogenesis}

The Affleck-Dine mechanism is an efficient mechanism of baryogenesis
in supersymmetric models \cite{Affleck:1984fy,Enqvist:2002gh}. In
fact, it is too efficient and the produced baryon asymmetry, $n_b/s$,
is in general too large. However, additional entropy release by the
dilaton decay may significantly dilute the baryon asymmetry
\cite{Linde:1985jy,Campbell:1998yi} and we examine this possibility
here. 

It is known that the Q-ball formation occurs for many Affleck-Dine flat
directions 
\cite{Enqvist:2002gh,Kusenko:1997si,Enqvist:1998en}. Whether 
Q-balls form or not depends on the shape of radiative correction to the flat
direction \cite{Enqvist:1998en,Enqvist:2000gq}.
In this paper we estimate the baryon asymmetry
providing that the Affleck-Dine field does not lead the
Q-ball formation.  For instance,
one example is a flat direction with large mixtures of stops
in the case of light gaugino masses, another is $H_u L$-direction
\cite{Enqvist:2000gq}.

\subsection{Original Affleck-Dine mechanism}

First, we investigate the originally proposed Affleck-Dine
mechanism with a flat potential up to  $\phi_{AD} \sim 1$
\cite{Affleck:1984fy}.
We consider the situation that there are the dilaton and the Affleck-
Dine
condensate in radiation dominated universe. As mentioned above, at the
moment $H = m_{\phi}$, the dilaton begins to oscillate with the initial
energy density $\rho_{\phi} \simeq 10^{-4}\rho_{\gamma}$, where
$\rho_{\gamma}$ is the energy density of the background
radiation. Then, on the other hand, the AD condensate is expected to
take a large expectation value, $\phi_{AD} \sim 1$, above which its
potential blows up exponentially.

The amplitude of the AD field and its energy density remains constant
while the Hubble parameter is larger than $m_{AD}$, where $m_{AD}$ is
the mass of the AD condensate. 
One should keep in mind, however, that though the energy 
density, $\rho_{AD}$, remains constant the baryon number
density decreases as $1/a^3$ if baryonic charge is conserved. So if
baryon charge is accumulated in ``kinetic'' motion of the phase of
$\phi_{AD}$ it would decrease as $1/a^3$. If, on the other hand, the
AD-field is frozen at the slope of not spherically symmetric potential
then baryonic charge of the AD-field is not conserved and after
$H < m_{AD}$ both radial and angular degrees of freedom would be
``defrosted'' and baryonic charge may be large.

As we noted above, 
when $H \simeq m_{3/2} (\simeq m_{AD})$, the AD field begins to
oscillate. Its energy density at that moment becomes comparable to
that of the radiation, while the energy density of the dilaton is
estimated as
\begin{eqnarray}
\left.\rho_{\phi}\right|_{H=m_{3/2}} =
\left(\frac{m_{\phi}}{m_{3/2}}\right)^{1/2}\left.\frac{\rho_{\phi}}{\rho
_{\gamma}}\right|_{H=m_{\phi}}\left.\rho_{\gamma}\right|_{H=m_{3/2}} 
 \simeq 10^{-3}\left.\rho_{\gamma}\right|_{H=m_{3/2}} ,
\end{eqnarray}
for the initial energy density $\rho_{\phi} = 10^{-4} \rho_{\gamma}$
and $m_{\phi} \simeq 10^2 m_{AD}$.
After that the universe becomes dominated by the oscillating AD
condensate and enters into approximately matter dominated regime (see
below). 

The energy density of the condensate and its baryon number density are 
given
respectively by the expressions:
\begin{eqnarray}
\rho_{AD} = m_{AD}^2 \phi_{AD}^2,\ ~~~~~~~~~~\ n_b =\kappa m_{AD}\phi_{
AD}^2 ,
\label{rhobad}
\end{eqnarray}
where $\kappa = n_b/n_{AD} <1$ is a numerical coefficient and $n_{AD}$ 
is the number density of the AD field. 

The rate of evaporation of the condensate, given by the decay width
of the AD field into fermions, $\Gamma_{AD} = C m_{AD}$ with  $C=
0.1-0.01$, is quite large. When the Hubble parameter becomes smaller
than $\Gamma_{AD}$, thermal equilibrium would be established rather
soon. 
However, the condensate would evaporate very slowly and disappear much
later~\cite{Dolgov:sy}. The low evaporation rate 
is related to a large baryonic charge and
relatively small energy density of the condensate. Below we will find
the temperature and the moment of the condensate evaporation repeating
the arguments of ref.~\cite{Dolgov:sy}. Let us assume that the
condensate evaporated immediately when $H= \Gamma_{AD}$ producing
plasma of relativistic particles with temperature $T_{AD}$ and chemical
potential $\mu_{AD}$. The temperature can be estimated as $T_{AD}
\simeq \rho_{AD}^{1/4}$ and since $T_{AD}\gg m_{AD}$ the chemical
potential is given by
\begin{eqnarray}
\mu_{AD} \simeq \frac{n_b }{T_{AD}^2} 
= \kappa \phi_{AD} \gg m_{AD} ,
\label{muad}
\end{eqnarray}
if $\kappa$ is not very small.
On the other hand, chemical potential of bosons cannot exceed their
mass. It means that instantaneous evaporation of the condensate is
impossible. The process of evaporation proceeds rather slowly with an
almost constant temperature of the created relativistic plasma. During
the process of evaporation the energy density of the latter was small in
comparison with the energy density of the condensate, except for the
final stage when the condensate disappeared. 

The cosmological baryon number density and energy densities are given by
 the 
equilibrium expressions:
\begin{eqnarray}
\frac{n_{b,\tot}}{T^3} &=& 
\frac{2N_f N_c B_q}{6\pi^2} \left( \xi_q^3 +\pi^2 \xi_q
\right) + \frac{1}{2\pi^2} \int_0^\infty d\eta \eta^2 \left[
\frac{1}{\exp (\epsilon -\xi ) -1} - \frac{1}{\exp (\epsilon +\xi ) -1}
\right] + B_c \label{Bt} \\
\frac{\rho_{\tot}}{T^4} &=& 
\frac{\pi^2 g_*}{ 30} +\frac{7}{8}\frac{2N_f\,(N_c+1)\pi^2}{15} \left[ 1
+ \frac{30}{7}\left( \frac{\xi_q}{\pi}\right)^2 +\frac
{15}{7}\left( \frac{\xi_q}{\pi}\right)^4 \right] \nonumber \\
&& + \frac{1}{ 2\pi^2} \int_0^\infty d\eta \eta^2\epsilon  \left[
\frac{1}{\exp (\epsilon -\xi ) -1} + \frac{1}{\exp (\epsilon +\xi ) -1}
\right] + \rho_c  
\label{rhot}
\end{eqnarray}
where $\eta \equiv p/T$ is dimensionless momenta, 
$\epsilon \equiv \sqrt{ \eta^2 + m_{AD}^2/T^2}$,
$\xi \equiv \mu_{AD} /T$ and $\xi_q \equiv \mu_q /T$
are dimensionless chemical potential of the AD field and quarks,
respectively.   $N_f =6$
and $N_c =3$ are the numbers of flavors and colors and factor 2 came    
 from
 counting spin states, $B_q = 1/3$ is
the baryonic charge of quarks while the baryonic charge of the
AD field is assumed to be 1, and $B_c$ and $\rho_c$ are baryon number 
density
and energy density of the condensate normalized to $T^3$ and
$T^4$, respectively. The first term in $\rho_\tot$ includes energy
density of light particles with zero charge asymmetry and $g_*$ is
the number of their species. 
The second term includes the contribution from leptons with the same
chemical potential as quarks - it is given by $(N_c+1)$.

For definiteness, let us assume 
that the AD field decays into the channel $\phi_{AD}
\rightarrow 3q+ l$ and taking into account that the sum of baryonic
and leptonic charges is 
conserved\footnote{One may wonder if this 
assumption is inappropriate because the
baryon asymmetry created in this channel would be washed out by
anomalous electroweak processes \cite{KRS}.  As will be seen later,
however, we can avoid this difficulty because in most cases of our
interest the AD condensate
evaporates at a lower temperature when these anomalous
processes are no longer effective.
}, so that $B-L=0$, we find
\begin{eqnarray}
\mu_q = \mu_l = \frac{\mu_{AD}}{4} .
\label{muq-mul}
\end{eqnarray}
Before complete evaporation of the condensate, the chemical potential of
AD-field remains constant and equal to its maximum allowed value
$m_{AD}$. Thus the only unknowns in these expressions are the
temperature and the amplitude of the field in the condensate. According 
to
Eq.(\ref{rhobad}), from Eqs.\ (\ref{Bt}) and (\ref{rhot}) we obtain
\begin{eqnarray}
\frac{B_c}{\rho_{c}} = \kappa \frac{T}{m_{AD}} .
\label{bc/rhoc}
\end{eqnarray}
The same relation was true for the initial values of $n_{b,\tot}/T^3$ 
and
$\rho_{\tot}/T^4$. 
Assuming that the ratio $n_{b,\tot}/\rho_{\tot}$ remains the
same during almost all process of evaporation, though it is not exactly 
so,  
we can exclude $B_c, n_{b,\tot}, \rho_c$ and $\rho_{\tot}$ from 
expressions 
 (\ref{Bt}) and  (\ref{rhot}) and find one equation that permits
to calculate the plasma temperature in presence of evaporating
condensate as a function of the baryonic charge fraction in
the initial condensate, $\kappa$. We find
\begin{eqnarray}
m_{AD} /T \simeq 20,\ ~~~~~~~~~~\ \textrm{for}\ ~~~~~~~~~~\  \kappa
=1, \\
m_{AD}/T \simeq 2,\ ~~~~~~~~~~\ \textrm{for}\ ~~~~~~~~~~\ \kappa =0.1.
\label{m/T}
\end{eqnarray}

Exact solution of the problem demands much more complicated study of
the evolution of energy density according to the equation
$\dot{\rho} = -3H(\rho + P)$, while the evolution of baryonic charge
density is determined by the conservation of baryonic charge which is
assumed to be true at the stage under consideration and thus  
$n_{b,\tot} \propto a^{-3}$.
The temperature of plasma found in this way would not be much 
different from the approximate expressions presented above.

Using the above-calculated plasma temperature (\ref{m/T}) we find that
at the moment of condensate evaporation (when $\mu_{AD} = m_{AD}$)
the cosmological energy and baryon number densities of the created
relativistic plasma are given by 
\begin{eqnarray}
\rho_p &\simeq & 1000 T^4, \label{rhop} \\
n_b &\simeq & 50 T^3, \label{bp}
\end{eqnarray}
for $\kappa =1$, and 
\begin{eqnarray}
\rho_p &\simeq & 70 T^4, \label{rhop2} \\
n_b &\simeq & 1.75 T^3, \label{bp2}
\end{eqnarray}
for $\kappa = 0.1$.

We see that a large baryon asymmetry prevents from fast condensate
evaporation, though the interaction rate could be much larger than the
expansion rate. From Eq.~(\ref{bp}) or (\ref{bp2}) and the baryon number
 conservation
\begin{equation}
n_b = \kappa m_{AD}\left(\frac{a_{AD}}{a(t)}\right)^3 
\lmk\phi_{AD}|_{H=m_{AD}}\rmk^2 ,
\end{equation}
we find 
\begin{eqnarray}
\lmk\frac{a_{\ev}}{a_{AD}}\rmk^3 &=& \kappa \frac{m_{AD}}{n_b}
\lmk\phi_{AD}|_{H=m_{AD}}\rmk^2
 \label{ze} \\
& \simeq &
160\, \lmk\frac{\phi_{AD}|_{H=m_{AD}}}{m_{AD}}\rmk^{2}  
 \simeq 10^{33} ,\ ~~~~~~~~~~\ \textrm{for}\ ~~~~~~~~~~\  \kappa=1,  \\
& \simeq &
0.46\,\lmk\frac{\phi_{AD}|_{H=m_{AD}}}{m_{AD}}\rmk^{2}
\simeq 3\times 10^{30} ,\ ~~~~~~~~~~\ \textrm{for}\ ~~~~~~~~~~\  \kappa=
0.1,
\end{eqnarray}
where $a_{AD}$ and $a_{\ev}$ are the value of the scale factor at the
moment $H=m_{AD}$ and that at the evaporation of the AD field, 
respectively.
Then at the evaporation the Hubble parameter and the baryon-to-entropy
ratio are respectively given by 
\begin{eqnarray}
H_{\ev} = (\rho_p/3)^{1/2} \simeq  \left\{ \begin{array}{ll}
2\times 10^{-14}\,\,\textrm{GeV} & \mbox{{for $\kappa =1$}},\\
5\times 10^{-13}\,\,\textrm{GeV} & \mbox{{for $\kappa =0.1$}}
 \end{array}\right.
\label{Hc}
\end{eqnarray}
\begin{eqnarray}
\left.\frac{n_b}{s}\right|_{\ev} &\simeq& 1 ,\ ~~~~~~~~~~\ \textrm{for}\
 ~~~~~~~~~~\  \kappa
=1, \\
\left.\frac{n_b}{s}\right|_{\ev} &\simeq& 0.04 ,\ ~~~~~~~~~~\ \textrm{
for}\ ~~~~~~~~~~\  \kappa
=0.1, 
\end{eqnarray}
from Eqs.\ (\ref{rhop}), (\ref{bp}), (\ref{rhop2}) and (\ref{bp2}).

Now we have to calculate the ratio of the baryon asymmetry to the
entropy of the plasma after thermalization of the products of dilaton
decay. Initially, at the moment of evaporation of AD-condensate the 
energy density of the dilaton is roughly $10^{-3}$ with respect to the
energy density of plasma. The latter is dominated by chemical potential
$\mu = m_{AD}>T$. When the universe expanded by the factor
$\left.\rho_{AD}/\rho_\phi \right|_\ev\equiv 
a_{\textrm{eq}}/a_{\ev} \simeq 10^3$ the dilaton starts to 
dominate and the relativistic expansion regime turns into matter 
dominated
one at $a=a_{\textrm{eq}}$. 
To the moment of the dilaton decay the energy density of the dilaton
becomes larger than the energy density of the plasma formed by the 
evaporation 
of the AD-condensate by the factor $a_{\textrm{d}}/a_{\textrm{eq}} =
(H_{\textrm{eq}}/H_{\textrm{d}})^{2/3}$, where $a_{\textrm{d}}$ and
$H_{\textrm{d}}$ are the scale factor and the Hubble parameter at the
time of the dilaton decay. Keeping in
mind that $H_{\textrm{eq}} = (a_{\textrm{eq}}/a_{\ev})^2 H_{\ev} =
10^{-6} H_{\ev}$,
we obtain the dilution factor by the dilaton decay
\begin{eqnarray}
\Delta = \left.\left( {\rho_\phi \over \rho_{AD}}
\right)^{3/4}\right|_{\textrm{d}} = 
\left( {H_{\ev} a_{\ev}^2 \over H_{\textrm{d}} a_{\textrm{eq}}^2 }\right)
^{1/2}=
\left.{\rho_{\phi} \over \rho_{AD}}\right|_{\ev}\,\left( {H_{\ev} \over 
H_{\textrm{d}}}\right)^{1/2}
\label{delta1}
\end{eqnarray}
Thus for $\kappa = 1$ the dilution factor is only 14, while for $\kappa 
= 0.1$
it is 70.

Finally we find that the baryon asymmetry after the dilaton decay
is given by
\begin{eqnarray}
\frac{n_b}{s} = \left.\frac{n_b}{s}\right|_{\ev}\frac{1}{\Delta} 
 \simeq 0.1-0.001.
\end{eqnarray}
Thus in this model the dilution of originally produced asymmetry
from the decay of AD field is not sufficient.

\subsection{Affleck-Dine mechanism with non-renormalizable potential}

As we have seen, additional entropy production due to the dilaton
decay is too small to dilute the baryon asymmetry generated in the
original Affleck-Dine scenario. Therefore it is necessary to suppress 
the generated baryon asymmetry. The presence of the non-renormalizable
terms can reduce the expectation value of the AD field during
inflation. As a result, the magnitude of the baryon asymmetry can be
suppressed. Hence we introduce the following non-renormalizable term 
in the superpotential to lift the
Affleck-Dine flat direction as a cure to regulate the baryon asymmetry
\cite{Ng:1988yn,Dine:1995kz}. 
\begin{equation}
W = \frac{\lambda}{n M^{n-3}}\phi_{AD}^n,
\end{equation}
where $M$ is some large mass scale.

The potential for the AD field in the inflaton-dominated stage reads
\begin{equation}
V(\phi_{AD}) = -c_1H^2|\phi_{AD}|^2 + \left(\frac{c_2\lambda H\phi_{AD}^
n}{n
M^{n-3}} + \hc \right) + |\lambda|^2\frac{|\phi_{AD}|^{2n-2}}{M^{2n-6}} ,
\end{equation}
where $c_1$ and $c_2$ are constants of order unity. The first and the
second terms are soft terms which arise from the supersymmetry breaking
effect due to the vacuum energy of the inflaton. The minimum of the
potential is reached at 
\begin{equation}
|\phi_{AD}| \simeq \left(\frac{H
M^{n-3}}{\lambda}\right)^{1/(n-2)} \label{min}.
\end{equation}
During inflation, the AD field takes the expectation value $|\phi_{AD}|
\simeq (H_{\textrm{inf}}M^{n-3}/\lambda)^{1/(n-2)}$, where
$H_{\textrm{inf}}$ is the Hubble parameter during inflation. After 
inflation, it also traces the instantaneous minimum, Eq.\
(\ref{min}), until the potential is modified and the field becomes
unstable there.  The AD field starts oscillation when its effective mass
becomes larger than the Hubble parameter.  There are three possible
contributions to trigger this oscillation: the low
energy supersymmetry breaking terms, the thermal mass term from
a one-loop effect \cite{Allahverdi:2000zd}, and the thermal effect at
the two-loop order \cite{Anisimov:2000wx}.


First we consider the case that the low energy supersymmetry breaking
terms are most important and that these thermal effects are negligible.
When $H \sim m_{3/2}$, the low energy supersymmetry breaking terms
appear and the potential for the AD field becomes
\begin{equation}
V(\phi_{AD}) = m_{AD}^2|\phi_{AD}|^2 + \left(\frac{Am_{3/2}\phi_{AD}^n}{
n
M^{n-3}} + \hc \right) + |\lambda|^2\frac{|\phi_{AD}|^{2n-2}}{M^{2n-6}} ,
\end{equation}
where $A$ is a constant of order unity, and the ratio of the energy
density of the AD field $\rho_{AD}$ to that of the inflaton
$\rho_I$ is given by
\begin{equation}
\frac{\rho_{AD}}{\rho_I} \simeq
\left(\frac{m_{3/2}M^{n-3}}{\lambda}\right)^{2/(n-2)} ,\label{rho-ad-i}
\end{equation}
up to numerical coefficients depending on $c_1,~c_2$ and $A$.
For example, the typical value for $n = 4$ is $\rho_{AD}/\rho_I \simeq
10^{-16}(M/\lambda)$.

Let us first assume that the inflaton decayed  at $H=m_{AD}
(\simeq m_{3/2})$ with the decay rate $\Gamma_I=m_{AD}$ and after that
the universe was dominated by radiation. Note that the
corresponding reheating temperature is $T_R \simeq \sqrt{\Gamma_I}
\simeq \sqrt{m_{AD}} \simeq 10^{10}$ GeV.  
Then the evaporation of the AD
condensate into relativistic plasma would be different from the
evaporation into cold plasma considered in the previous
subsection. Due to interaction with plasma the products of the
evaporation acquire much larger temperature than in the case of the
evaporation into  vacuum. Since the energy density of the condensate
is negligible in comparison with the total energy density of the
plasma, the temperature of the latter drops in the usual way,
$T\propto 1/a$, in contrast to the previously considered case when $T=
const$. Since the temperature of the plasma is high, $T\gg m_{AD}$, the
baryon number density is $n_b = B_c T^3 +C_B T^2 \mu$ where $C_B\sim 1$
is a constant coefficient, $\mu\leq m_{AD}$ is the value of the
chemical potential, and we have neglected terms of the order of
$\mu^3$. Since $n_b \propto a^{-3}$, the ratio of $a_{\ev}$ to
$a_{AD}$ is
\begin{eqnarray}
\frac{a_{\ev}}{a_{AD}} &=& \frac{n_b|_{H=m_{AD}}}{m_{AD} T_R^2} \\
&\simeq&
10\left(\frac{m_{AD}}{1\textrm{TeV}}\right)\left(\frac{10^{10}\textrm{GeV}}
{T_R}\right)^2\left(\frac{M}{\lambda}\right)\,
~~~~~~~~~~~~ \textrm{for} \,~~~~~~~~~~~~~~~\ n=4,
%
\label{zev}
\end{eqnarray}
(compare to Eq.\ (\ref{ze})). 
Here we took for the initial value of the baryonic charge density
$n_b|_{H=m_{AD}} = \kappa m_{AD}\phi_{AD}^2$ with $\kappa \sim 1$. 
For $n=4$, $a_{\ev}/a_{AD}$ becomes of order $10$,
and we find that the condensate would evaporate soon.

To be more precise, however, we must take into account that the
interaction rate of the condensate is $\Gamma_{AD} = (0.1-0.01)m_{AD}$
and the evaporation cannot start before $H=\Gamma_{AD}$. At that
moment the plasma temperature would be smaller by the factor 
$(m_{AD}/\Gamma_{AD})^2 = 10^2-10^4 $ and the baryon number
density of the condensate would be smaller by
$(m_{AD}/\Gamma_{AD})^6$. Correspondingly the red-shift of the end of 
evaporation should be
shifted by a factor $(m_{AD}/\Gamma_{AD})^2 $ with respect to the 
beginning of
evaporation and it means that it would remain the same with respect
to the initial moment $H=m_{AD}$.

As we have already noted, for $n=4$ the condensate decays quickly and 
the baryon number density produced in the decay is diluted by the
plasma created by the inflaton decay as 
\begin{eqnarray}
\frac{n_b}{s} \simeq \frac{T_R}{m_{AD}}\frac{\rho_{AD}}{\rho_I} 
= 10^{-9}\left(\frac{T_R}{10^{10}\textrm{GeV}}\right)
\left(\frac{1\textrm{TeV}}{m_{AD}}\right)
\left(\frac{M}{\lambda}\right) \label{del4}.
\end{eqnarray}
where $T_R$ is the reheating temperature of the inflaton and we used
the estimate of Eq.\ (\ref{rho-ad-i}). 
The result does not depend upon the moment of the decay of
AD-condensate, since its energy density remains
sub-dominant. 
If an additional dilution by the dilaton and the early oscillation by
a thermal effect \cite{Allahverdi:2000zd} are operative, the
baryon asymmetry become even smaller than the observed one and the
$n=4$ model cannot explain the observed baryon asymmetry. Hence we
must consider the flat direction with $n > 4$. 

Hereafter, we study the AD fields with $n > 4$ including the thermal
effect. For the AD fields with $n > 4$, the relevant thermal effect comes from
the running of the gauge coupling constant \cite{Anisimov:2000wx}
rather than the thermal plasma effect \cite{Allahverdi:2000zd}.    
The potential for the AD field in the inflaton-dominated stage reads
\begin{eqnarray}
V(\phi_{AD}) &=& (-c_1H^2 + m_{AD}^2)|\phi_{AD}|^2 + \alpha T^4
\ln\left(\frac{|\phi_{AD}|^2}{T^2}\right) \nonumber \\
 & & + \left(\frac{c_2\lambda H\phi_{AD}^n}{n M^{n-3}} +
\frac{Am_{3/2}\phi_{AD}^n}{n M^{n-3}} + \hc \right) +
|\lambda|^2\frac{|\phi_{AD}|^{2n-2}}{M^{2n-6}} ,
\label{pot-withT}
\end{eqnarray}
where the second term is the thermal effect at two loop order which
is pointed out in \cite{Anisimov:2000wx} and $\alpha$ denotes the gauge
coupling.

If the effective mass of the AD field becomes comparable to the Hubble 
parameter when it is larger than the low energy supersymmetry breaking
scale,
\begin{eqnarray}
\alpha \frac{T^4}{|\phi_{AD}|^2} \simeq H^2 ~~ ( ~ > m_{AD}^2 ) ,
\label{begin:osc}
\end{eqnarray}
then AD field undergoes the early oscillation by the thermal
effect. During the oscillating inflaton dominated stage
($t<t_{\textrm{rh}}$), the
temperature of the plasma behaves as
\begin{eqnarray}
T \simeq T_R\left(\frac{a(t_{\textrm{rh}})}{a(t)}\right)^{3/8}
 \simeq T_R^{1/2}H^{1/4}.
\label{Temp}
\end{eqnarray}
From Eqs.\ (\ref{min}) and (\ref{Temp}), the effective mass term of
the AD field is rewritten as
\begin{equation}
\alpha\frac{T^4}{|\phi_{AD}|^2} \simeq \alpha
T_R^2\left(\frac{\lambda}{M^{n-3}}\right)^{2/(n-2)}H^{(n-4)/(n-2)}.
\label{thermal:mass}
\end{equation}
By comparing with Eq.\ (\ref{begin:osc}), when the AD
field begins to oscillate at $t\equiv t_{\os}$, 
the Hubble parameter is given by 
\begin{eqnarray}
H_{\os} \simeq
\left(\alpha\, T_R^2\right)^{(n-2)/n}
\left(\frac{\lambda}{M^{n-3}}\right)^{2/n}.
\label{Ho} 
\end{eqnarray}

From now on we concentrate on the case $n=6$, for which
\begin{eqnarray}
H_{\os} &\simeq& 
1\textrm{TeV}\left(\frac{\alpha}{10^{-2}}\right)^{2/3}
\left(\frac{T_R}{10^{8}\textrm{GeV}}\right)^{4/3}
\left(\frac{\lambda}{M^3}\right)^{1/3}.
\end{eqnarray}
Thus we find that the AD field begins to oscillate at
 $H \gtrsim m_{AD}$ due to the thermal term 
if $T_R \gtrsim 10^8$  GeV for $M = 1$,
  and if  $T_R \gtrsim 10^6$ GeV for
 $M = 10^{-2}$, respectively.
During the early oscillation driven by the thermal term, 
the amplitude of the AD field decreases as
\begin{equation} 
|\phi_{AD}(t)| = |\phi_{AD}|_{t_{\os}}\left(\frac{a_{\os}}{a(t)}\right)^{9/4} ,
\label{AD-dec}
\end{equation}
where $a_{\os}$ denotes the scale factor at the beginning of
the oscillation. The analytic derivation of Eq.\ (\ref{AD-dec}) is
shown in Appendix and we have confirmed this result by the numerical
calculation. It also agrees with the analysis in \cite{Fujii:2001dn}. 
Using Eqs. (\ref{min}), (\ref{Temp}) and
(\ref{AD-dec}), we obtain the ratio of the amplitude of the AD field
to temperature of plasma as
\begin{equation} 
\frac{|\phi_{AD}|}{T} = \left(\frac{M^3}{T_R^2\lambda}\right)^{1/4}
\left(\frac{a_{\os}}{a(t)}\right)^{15/8} .
\end{equation}
We consider the evaporation rate of the AD field. The condition
that the particles coupled to the AD field are 
light enough to exist as much as radiation
reads $h|\phi_{AD}| < T$, where $h$ is the corresponding coupling
constant.
Following \cite{Allahverdi:2000zd}
let us adopt the scattering rate of the AD field  $\Gamma \sim
h^4 T$ in this situation as the rate of its evaporation.
Estimating  $h|\phi_{AD}| /T$ and $\Gamma$ at the
reheating time, we find
\begin{equation} 
\left.\frac{h|\phi_{AD}|}{T}\right|_{t_{\textrm{rh}}} 
\simeq 10^{-1}\left(\frac{T_R}{10^{16}\textrm{GeV}}\right)^{1/3}
\left(\frac{h}{10^{-2}}\right)\left(\frac{10^{-2}}{\alpha}\right)^{5/6}\left(\frac{M^3}{\lambda}\right)^{2/3} ,
\label{exp-value}
\end{equation}
and
\begin{eqnarray} 
\frac{\Gamma}{H} \simeq \frac{h^4}{T_R} \simeq
\left(\frac{10^{10}\textrm{GeV}}{T_R}\right)\left(\frac{h}{10^{-2}}\right)^4 .
\label{Gamma}
\end{eqnarray} 
Eq. (\ref{exp-value}) shows that the particles coupled to the AD field
are thermally excited and populated well before
 the reheating time for any reheating
temperature, $T_R \lesssim 10^{16}$GeV, while
 we find that the AD condensate can evaporate 
around the typical reheating time from Eq. (\ref{Gamma}). 

The baryon number density for the AD field $\phi_{AD}$ is given as
\begin{equation}
n_b = - i q (\phi_{AD}^* \dot{\phi}_{AD}-\dot{\phi}^*_{AD}\phi_{AD}),
\end{equation}
where $q$ is a baryonic charge for the AD field.

The baryon number density at $H=H_{\os}$ is estimated as
\begin{eqnarray}
\left.n_b\right|_{t_{\os}} = \left.\frac{4 q m_{3/2}}{3 H
M^3}\text{Im}(A\phi_{AD}^6)\right|_{t_{\os}} 
 = \frac{4 q \delta m_{3/2}}{3 \lambda}\left(\frac{H_{\os}
 M^3}{\lambda}\right)^{1/2} ,
\end{eqnarray}
where $\delta$ is a effective relative CP phase. The baryon-to-entropy
ratio at the reheating time is estimated as 
\begin{eqnarray}
\left.\frac{n_b}{s}\right|_{t_{\textrm{rh}}} = \frac{3T_R}{4}
\left.\frac{n_b}{\rho_I}\right|_{t_{\os}} 
 = \frac{q \delta
 m_{3/2}}{3\lambda}\frac{T_R}{H_{\os}^2}
\left(\frac{H_{\os}M^3}{\lambda}\right)^{1/2} .
\label{b/s:reheat}
\end{eqnarray}
Since the dilution factor by the dilaton decay is given by 
\begin{equation} 
\Delta = \frac{T_R}{10^4 T_D}
\left(\frac{\left.\rho_{\phi}/\rho_I\right|_{t_{\min}}}{10^{-4}}\right) ,
\label{dilute}
\end{equation}
from Eqs. (\ref{b/s:reheat}) and (\ref{dilute}), we obtain the final
baryon asymmetry,
\begin{eqnarray}
\frac{n_b}{s}
 &=& \left.\frac{n_b}{s}\right|_{t_{\textrm{rh}}}\frac{1}{\Delta} \nonumber \\
 &\simeq& 10^{-8}q \delta\, \left(\frac{m_{3/2}}{H_{\os}}\right)^{3/2}\left(\frac{M}{\lambda}\right)^{3/2}\left(\frac{T_D}{10^{-1}\textrm{GeV}}\right)
\left(\frac{1\textrm{TeV}}{m_{3/2}}\right)^{1/2}
\left(\frac{\left.\rho_{\phi}/\rho_I\right|_{t_{\min}}}{10^{-4}}\right)^{-1},
\end{eqnarray}
for $H_{\os} >m_{3/2}$.
This result can easily meet the observation if we take, for example, 
$H_{\os} \simeq 10^2 m_{3/2}$ and other factors to be of order of unity.
In this case, from Eq.\ (\ref{Gamma}), 
we find that for the present case the AD condensate
can evaporate before the inflaton decay is completed.

On the other hand, in the case  the reheating temperature is
so low that the thermal effect does not lead the early
oscillation, the reduction of the cut-off scale $M$ can lead to the
reasonable baryon asymmetry:
\begin{eqnarray}
\frac{n_b}{s} \simeq 10^{-11}q\delta 
\left(\frac{M}{10^{-2}\lambda}\right)^{3/2}
\left(\frac{T_D}{10^{-1}\textrm{GeV}}\right)
\left(\frac{1\textrm{TeV}}{m_{3/2}}\right)^{1/2}
\left(\frac{\left.\rho_{\phi}/\rho_I
\right|_{t_{\min}}}{10^{-4}}\right)^{-1} .
\end{eqnarray}
This expression implies that the cut-off scale $M$ should be around the GUT
scale $10^{-2}$ or $10^{16}$GeV
 for inflation models with a low reheating temperature,
$T_R\lesssim 10^{6}$GeV.
Furthermore we find that the final baryon asymmetry is independent of the
reheating temperature of inflation within the range.

In the present model, the supersymmetry breaking is caused by the F-term
of the dilaton. Therefore, when the dilaton decays, it can decay into
gravitinos through their mass term and this process could lead to 
overproduction of the gravitinos. The constraint derived in
\cite{Hashimoto:1998mu} to avoid the overproduction
 is $m_{\phi} \gtrsim 100$ TeV. The mass of the 
dilaton, (\ref{mass}), in the model considered here is  in the allowed
region. 

\section{Conclusion}

In this paper, we have studied the Affleck-Dine baryogenesis in the
framework of the string cosmology. In string models, the dilaton is
ubiquitous and does not have any potential perturbatively. We
adopted the non-perturbatively induced potential of the dilaton via
the gaugino condensation in the hidden gauge sector. Then we set
phenomenologically desired
values for the
gravitino mass and the VEV of the dilaton.

The attractive mechanism to stabilize the dilaton at the desired minimum
was proposed by Barreiro \textit{et al}. \cite{Barreiro:1998aj}. 
They did not estimate
the energy density of the oscillating dilaton.  It is estimated in
the presented paper where we have 
found $\rho_{\phi} \simeq 10^{-4}\rho$ 
at $H = m_{\phi}$. This energy transforms into the radiation after the 
decay of the dilaton before nucleosynthesis because 
the mass $m_{\phi} \simeq 10^2$ TeV is sufficiently high.

We have discussed cosmological baryogenesis in this model. 
In the above-mentioned cosmological history 
with the entropy production, the Affleck-Dine baryogenesis might be
the only workable mechanism for baryogenesis. We have investigated the
Affleck-Dine baryogenesis with and without non-renormalizable terms.
We have shown that while the original Affleck-Dine scenario produces too
much baryon asymmetry even if there is the dilution by the dilaton
decay, the model with $n = 6$ non-renormalizable terms can lead to the
appropriate baryon asymmetry.

\acknowledgements
{
A.D. is grateful to YITP for the hospitality during period when this
work was done. O.S. and J.Y. are grateful to John Ellis for useful 
discussion. We are grateful to Masaaki Fujii for informative comments.
The work of J.Y. is partially supported by JSPS Grant-in-Aid for
Scientific Research 13640285.
}

\appendix
\section*{}

In this appendix, we derive Eq.\ (\ref{AD-dec}). Although the similar 
discussion can be found in the literature \cite{deGouvea:1997tn}, we
derive it for completeness. 

We consider the evolution of the AD field after the beginning of
oscillation induced by the thermal effect at the two-loop level.
Then the AD field obeys the equation of motion
\begin{equation}
\ddot{\phi}_{AD}+3H\dot{\phi}_{AD}+\alpha\frac{T^4}{\phi^*_{AD}} = 0 .
\label{ADeom}
\end{equation}
By decomposing $\phi_{AD}$ into
\begin{eqnarray}
\phi_{AD} = |\phi_{AD}|e^{i \theta} \equiv \Phi e^{i \theta} ,
\end{eqnarray}
Eq.\ (\ref{ADeom}) is reduced to the following equations
\begin{eqnarray}
&& \ddot{\Phi}+3H\dot{\Phi}-\dot{\theta}^2\Phi+\alpha\frac{T^4}{\Phi} = 0, 
\label{Xeom}\\
&& (a^3 \dot{\theta}\Phi^2)\dot{} = 0 .\label{theta}
\end{eqnarray}
The second equation (\ref{theta}) is interpreted as the conservation of 
the angular momentum which corresponds to the baryon number density
and rewritten as
\begin{eqnarray}
\dot{\theta}\Phi^2 =
\left.\dot{\theta}\Phi^2\right|_{t_{\os}}\left(\frac{a_{\os}}{a(t)}\right)^3
\equiv m\Phi_0^2\left(\frac{a_{\os}}{a(t)}\right)^3 ,
\end{eqnarray}
where $\Phi_0$ represents the initial amplitude of the AD field and $m$ 
means the initial angular velocity of the order of $m_{3/2}$ for $n =6$.
By eliminating $\dot{\theta}$ in Eqs.\ (\ref{Xeom}) and (\ref{theta}),
we obtain
\begin{equation}
\ddot{\Phi}+3H\dot{\Phi}-m^2\left(\frac{a_{\os}}{a(t)}\right)^6\left(\frac{\Phi_0}{\Phi}\right)^4\Phi+\alpha\frac{T^4}{\Phi}
= 0 .
\label{CHIeom}
\end{equation}
Multiplied by $\dot{\Phi}$ and using $T^4 \propto a^{-3/2}$, Eq.\
(\ref{CHIeom}) yields
\begin{eqnarray}
&&\frac{d}{d
t}\left[\dot{\Phi}^2+m^2\left(\frac{a_{\os}}{a(t)}\right)^6\frac{\Phi_0^4}{\Phi^2}+\alpha
T^4\ln\frac{\Phi^2}{T^2} \right] \nonumber\\
&& = -6H\left[\dot{\Phi}^2+m^2\left(\frac{a_{\os}}{a(t)}\right)^6\frac{\Phi_0^4}{\Phi^2}
\right] -\frac{3}{2}H\alpha T^4\ln\frac{\Phi^2}{T^2} +
\frac{3}{4}H\alpha T^4 .
\label{energy}
\end{eqnarray}

On the other hand, multiplying Eq.\ (\ref{CHIeom}) by $\Phi$, we obtain
\begin{equation}
\frac{1}{a^3}(\Phi a^3 \dot{\Phi})\dot{}
-\dot{\Phi}^2-m^2\left(\frac{a_{\os}}{a(t)}\right)^6\frac{\Phi_0^4}{\Phi^2}+\alpha 
T^4 = 0.
\end{equation}
By taking the time average over the time scale of the cosmic
expansion, we obtain the cosmic virial theorem 
\begin{equation}
\left\langle\dot{\Phi}^2+m^2\left(\frac{a_{\os}}{a(t)}\right)^6\frac{\Phi_0^4}{\Phi^2} \right\rangle = \langle\alpha T^4\rangle ,\label{cvt}
\end{equation}
where $\langle...\rangle $ denotes the time average. Moreover, since the
second term which represents the centrifugal force becomes efficient
around only $\Phi \approx 0$, Eq.\ (\ref{cvt}) could be rewritten as
\begin{equation}
\langle\dot{\Phi}^2\rangle = \langle\alpha T^4\rangle .
\label{cvt2}
\end{equation}

From Eqs.\ (\ref{energy}) and (\ref{cvt2}), we obtain
\begin{equation}
\frac{d}{d t}\left(1 + \ln\frac{\Phi^2}{T^2}\right) = -\frac{15}{4}H .
\end{equation}
Since we know $T \propto a^{-3/8}$, hence, we find 
\begin{equation}
\Phi \propto a^{-9/4} ,
\end{equation}
for $\Phi \gtrsim T$.


\begin{figure}
 \begin{center}
\centerline{\epsfxsize=1.0\textwidth\epsfbox{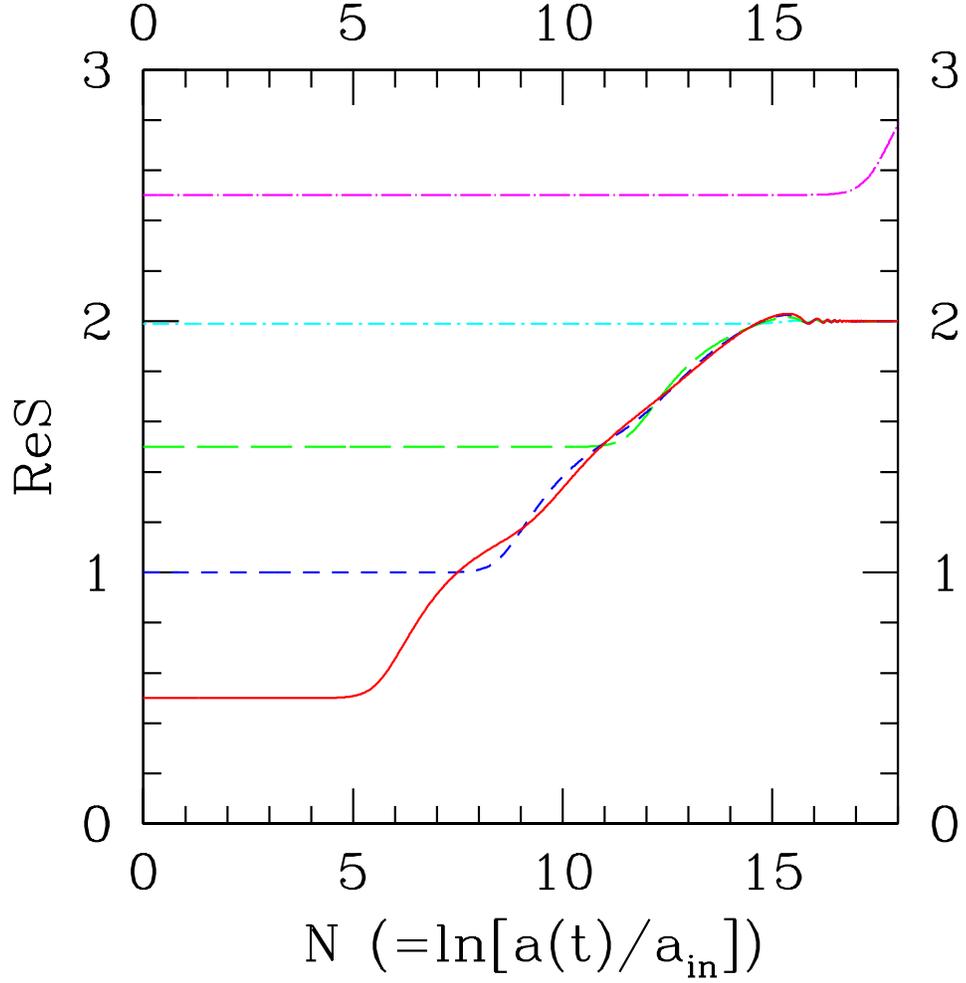}}
 \end{center}
\caption{
Evolution of Re $S$ as a function of $N$ for various initial
conditions with $S_{\min}=2$ in case of $\gamma = 4/3$. We set the
initial values of the velocity and the Hubble expansion rate as $d
{\rm Re}S/dt|_{0} = 0$, and $H_{0} =1$, respectively.
}
\label{fig1}
\end{figure}

\newpage

\begin{figure}
 \begin{center}
 \centerline{\epsfxsize=0.6\textwidth\epsfbox{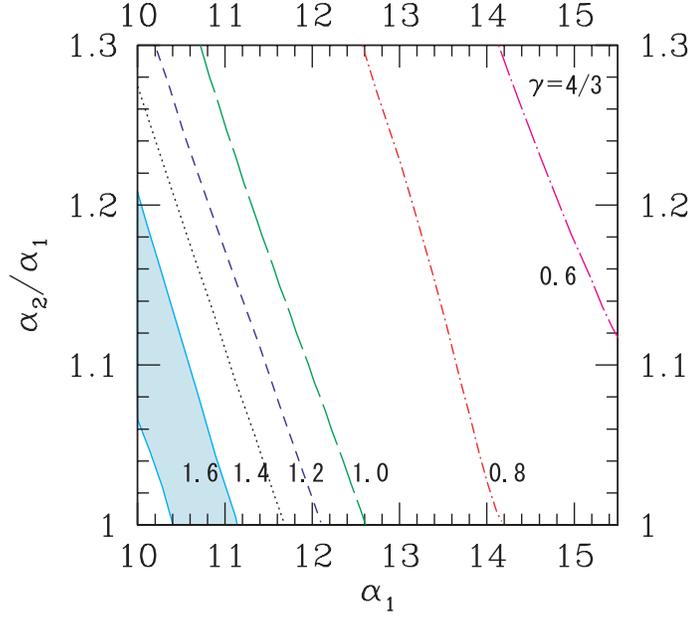}}
 \end{center}
\caption{
Energy density of the dilaton at the beginning of the oscillations for
various model parameters. The number associated with each contour line
represents the value of $\rho_{\phi}$ normalized by $10^{-4}\rho$.
Here we adopt $\gamma = 4/3$.
}
\label{fig2}
\end{figure}

\begin{figure}
 \begin{center}
     \centerline{\epsfxsize=0.6\textwidth\epsfbox{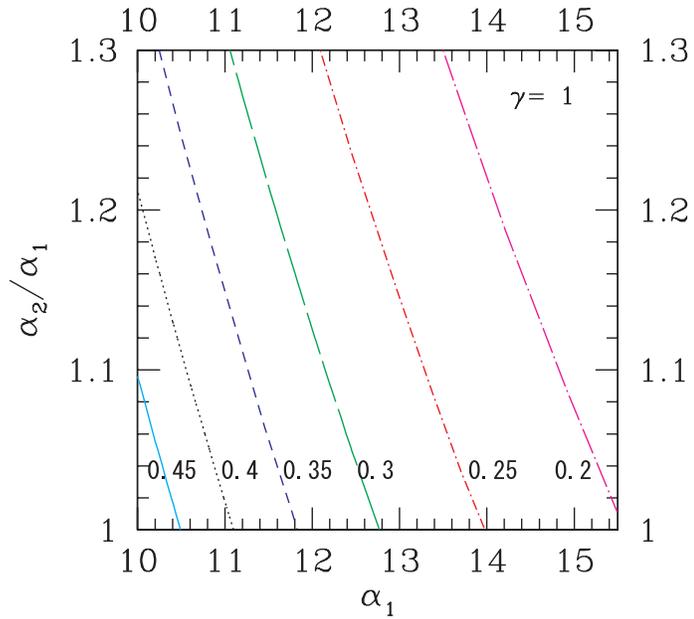}}
 \end{center}
\caption{
Same as Fig.~2 except for $\gamma$ = 1.
}
\label{fig3}
\end{figure}
\end{document}